\documentclass[hyper]{JHEP3}

\input{epsf}
\usepackage{epsfig}
\usepackage{amssymb}
\usepackage{amsfonts}
\usepackage{amsbsy}
\usepackage[all,2cell]{xy}

\usepackage{amsmath}

\usepackage{amssymb,amscd}
\usepackage{mathrsfs}
\usepackage{amsmath,amsthm}

\def\be{\begin{equation}}
\def\ee{\end{equation}}

\def\ID{\mathbb{D}}

\def\CO {{\cal O}}

\def\half{\frac{1}{2}}

\def\one{{\hbox{ 1\kern-.8mm l}}}

\def\vol{{\rm vol\,}}
\def\p{\partial}

\def\be{\bar{e}}

\def\half{\frac{1}{2}}

\def\half{\frac{1}{2}}

\title{A Shifted View of Fundamental Physics }

\author{Michael    Atiyah and  Gregory W. Moore  }

\abstract{We speculate on the role of relativistic versions of delayed differential equations
in fundamental physics. Relativistic invariance implies that we must consider both
advanced and retarded terms in the equations, so we refer to them as shifted equations.
The shifted Dirac   equation has some novel properties. A tentative formulation
of shifted Einstein-Maxwell equations naturally incorporates a small but nonzero
cosmological constant. }

\begin{document}

\section{Prologue by Michael   Atiyah }

 The past three or four decades have seen a remarkable fusion of
 theoretical physics with mathematics.  Much of the impetus for this is due to
 Is Singer whose 85th birthday is being celebrated at this meeting.
 He introduced me \footnote{This joint article is based on the lecture by the first author at Singer 85} to many of the physical ideas and instructed me in the areas of mathematics that had interacted strongly with physics in the first part of the 20th century.  These were differential geometry, flourishing in the wake of Einstein's theory of general relativity and functional analysis, which provided the rigorous background for quantum mechanics as laid down by von Neumann.

 By contrast my own mathematical background has
 centered round algebraic geometry and topology, the areas
  of mathematics which have played an increasingly important
   part in the developments in physics of the last quarter of the 20th century and beyond.

 I think Is and I were fortunate to be at
 the right place at the right time, working on the kind of
 mathematics which became the new focus of interest in theoretical
 physics of gauge theories and string theory.

 Looking to the past the great classical era began with
Newton, took a giant stride with Maxwell and culminated in Einstein's spectacular theory of general relativity.

 But then came quantum mechanics and quantum
 field theory with a totally new point of view, rather far removed from the geometric ideas of the past.

 The major problem of our time for physicists is
 how to combine the two great themes of GR, that governs the
 large scale universe, and QM that deals with the very small scale.

 At the present time we have string theory, or
perhaps ``M-theory", which is a beautiful rich mathematical
story and will certainly play an important role in the future of both
mathematics and physics.  Already the applications of these ideas to mathematics
have been spectacular.  To name just a few, we have

\begin{enumerate}
\item Results on the moduli spaces of Reimann surfaces.
\item The Jones polynomials of knots and their extension by Witten to ``quantum invariants" of 3-manifolds.
\item Donaldson theory of 4-manifolds and the subsequent emergence of Seiberg-Witten theory.
\item Mirror symmetry between holomorphic and symplectic geometry.
\end{enumerate}

Although string theory or M-theory are thought by many to be the ultimate theory combining QM and GR no-one   knows what M-theory really is.  String theory is recognized only as a perturbative theory, but the full theory is still a mystery (one of the roles of the letter M).

 Some claim that the final
theory is close at hand - we are almost there.  But perhaps this is misplaced
optimism and we await a new resolution based on radical new ideas.  There are,
after all, some major challenges posed by astronomical observations.

\begin{itemize}
  \item Dark Matter
  \item Dark Energy (with a very small cosmological constant)
\end{itemize}

 Moreover, the direct linkage between the rarified mathematics
of string theory and the world of experimental physics is, as yet,
 very slender.  A friend of mine, a retired Professor of Physics,
 commentating on a string theory lecture, said it was "pure poetry"!
 This can be taken both as criticism and as a tribute.  In the same way
 science fiction cannot compete with the modern mysteries of the quantum vacuum.

 But, if we need new ideas, where will they come from?  Youth is the traditional source of radical thoughts, but only a genius or a fool would risk their whole future career on the gamble of some revolutionary new point of view.  The weight of orthodoxy is too heavy to be challenged by a PhD student.

 So it is left to the older generation like me to speculate.  The same friend who likened string theory to poetry encouraged me to have wild ideas, saying "you have nothing to lose!" That is true, I have my PhD.  I do not need employment and all I can lose is a bit of my reputation.  But then allowances are made for old-age, as in the case of Einstein when he persistently refused to concede defeat in his battle with Niels Bohr.

 So my birthday present to Is is to tell him that we senior citizens can indulge in wild speculations!

\vfill\eject

\section{An Exploration}

The idea we\footnote{The work of G.M. is supported by DOE grant DE-FG02-96ER40959. He would
like to thank T. Banks and S. Thomas for discussions.}
 want to explore is the use of retarded (or advanced) differential equations in fundamental physics.
These equations, also known as
 ``functional differential equations,'' or ``delayed differential equations'' have been much studied by
engineers and mathematicians, but the applications in fundamental physics have been limited.
This idea has a number of different origins:

\begin{description}
\item[(i)] Such retarded differential equations occur in Feynman's thesis \cite{FeynmanThesis}.
\item[(ii)] In the introduction to Bjorken and Drell \cite{BD}  it is suggested that, if space-time at very small scales is ``granular," then one would have to use such equations.
\item[(iii)] Their use in the context of quantum mechanics has been advocated by C.K. Raju \cite{Raju}.
\end{description}

\noindent We will discuss the general idea briefly before going on to explain how to develop a more
scientific version.  This one of us worked on at an earlier stage as reported in the Solvay Conference
\cite{SolvayTalk}.

  Let us begin by looking at the simplified example of a linear retarded differential equation for a  function $x(t)$:
\begin{equation}\label{eq:ret}
\dot{x}(t)+kx(t-r)=0
\end{equation}
 where the positive number $r$ is the retardation parameter and $k$ is a constant. A 
 rescaling of the time variable shows that the equation really only depends on a single
 dimensionless parameter $\mu = kr$.  Moreover,
 the initial data for such an equation is an arbitrary function $g(t)$ over the interval $[0,r]$.
  Successive integration then allows us to extend the function for all $t \geq 0$, while successive differentiation (for smooth initial data $g(t)$) enables us to extend to negative $t$.  A second way to discuss the solutions is to note that the functions
 $x(t) = x_0 e^{- z t/r}$ solve \eqref{eq:ret} provided $z = \mu e^z$. The latter transcendental
 equation has an infinite set of roots tending to $z=\infty$. (In general, all the roots $z$ have
 nonzero real part, and hence the solutions have an unphysical divergence in the far past or future.)

From either approach,  we note that  equation    \eqref{eq:ret},
  like the equations of quantum mechanics, has an infinite-dimensional space of initial data:
   It can be taken to be the Hilbert space $L^2[0,r]$. We will take that as an encouraging sign and,
 without pursuing further the  parallel with quantum theory at present we can ask whether retarded differential equations make any sense in a relativistic framework where there is no distinguished time direction in which to retard. In fact this can be done and there is a natural and essentially unique way to carry this out.  The first observation is that the translation $t\rightarrow t-r$ has the infinitesimal generator $-r\frac{d}{dt}$, so that the translation is formally just $\exp(-r\frac{d}{dt})$.
 In Minkowski space we need a relativistically invariant version of $\frac{d}{dt}$, i.e. a relativistically invariant first order differential operator.  But this is just what Dirac was looking for when he invented the Dirac Operator $D$.  The important point is that $D$ acts not on scalar functions but on \textbf{spinor} fields.  Thus, we can write down a relativistic analogue of \eqref{eq:ret}  which is a retarded version of the usual Dirac equation
\begin{equation}\label{eq:ret-Dir}
\{i\hbar D-mc+ik \exp(-rD)\}\psi =0
\end{equation}
 where $D$ is the Dirac operator and $\psi$ is a spinor field.  Note that $r$ has dimensions of length
 $L$  whilst  $k$ (slightly different from that
appearing in \eqref{eq:ret}) now has the physical dimension $MLT^{-1}$.

  Both $k$ and $r$  are required to be real for physical reasons which will be clarified shortly (see
\eqref{eq:quant} and \eqref{eq:equiv-dev} below).  This appears to be rather a formal equation and one can question whether it makes any sense.  In fact \eqref{eq:ret-Dir} makes sense for all physical fields $\psi$, i.e. those which propagate at velocities less than the velocity of light.  Any such $\psi$ is a linear combination of plane-waves and the operator $\exp(-rD)$ applied to such a plane-wave component just retards it by $r$ in its own time-direction.
For waves which travel with velocity $c$ mathematical arguments based on continuity, or physical arguments using  clocks, require that there be no retardation.
 
 Although we have said that \eqref{eq:ret-Dir} is a retarded equation the fact that spinors have both positive and negative frequencies implies that it is also an advanced equation.  Perhaps we should use a neutral word such as ``shifted" instead of advanced or retarded. Having said that we may consider several variants of the shifted
Dirac equation where we replace

\begin{subequations}\label{eq:D-mod}
\begin{align}
D & \to    \ID_+ := D + \frac{k}{\hbar} e^{-r D} \label{eq:D-plus}  \\
D & \to  \ID_- := D - \frac{k}{\hbar} e^{r D} \label{eq:D-minus} \\
D & \to   \ID_s := D + \frac{k}{\hbar} \sinh(r D) \label{eq:D-s} \\
D & \to   \ID_c := D + i \frac{k}{\hbar} \cosh(rD) \label{eq:D-c}
\end{align}
\end{subequations}

For brevity we will focus on the modification \eqref{eq:D-plus} in what follows.

%
%\noindent We may regard equation \eqref{eq:ret-Dir}
%as a modification of the Dirac equation $i \hbar
%

\medskip

It is instructive to examine the plane-wave solutions of the modified Dirac equations
in Minkowski space.
We take $\psi = s(p) e^{-i p\cdot x/\hbar}$ where $s(p)$ is a constant spinor and
there is a dispersion relation  $p^2 = p_\mu p^\mu = E_0^2/c^2$, with $E_0> 0$
representing the inertial rest energy of a particle. Acting on such a wavefunction
$D \psi = \frac{E_0}{c} \gamma\cdot \hat p \psi $ where $\gamma\cdot \hat p$ squares to $1$.
Let $s_\pm(p)$ denote the eigenspinors.
The planewave solutions of the shifted Dirac equation
\begin{equation}\label{eq:mod-Dir}
(i \hbar \ID_+ - mc )\psi =0
\end{equation}
are then $ s_+(p)e^{-i p\cdot x/\hbar}$ and $s_-(p)e^{i p\cdot x/\hbar}$ provided
\begin{equation}\label{eq:ener-q}
\frac{E_0}{c}-mc+ik \exp(\dfrac{irE_0}{\hbar c})=0.
\end{equation}
Since the first two terms are real we derive a   quantization condition
\begin{equation}\label{eq:quant}
\dfrac{rE_0}{\hbar c}=(n+1/2)\pi \hspace{15pt} \textrm{with} \hspace{15pt}  n \hspace{15pt} \textrm{integral}
\end{equation}
 Note that, thanks to the half-integer shift,
  the value $r=0$ is excluded in \eqref{eq:quant}.\footnote{Given the speculative connections to
   quantum mechanics mentioned above it is natural to wonder if this is related to the zero point energy
   of the harmonic oscillator.} This is related to and explains the factor $i$ in \eqref{eq:ret-Dir} ($k$ being real).  Replacing $n$ by $n-1$ in \eqref{eq:quant} is equivalent to changing the sign of $k$ in \eqref{eq:ret-Dir}.  The two signs of $k$ are on an equal footing and so we should consider both.
 A similar discussion with a massless dispersion relation, i.e. $E_0=0$, shows that there
 are no solutions: It is not possible to retard a massless stable fermion.

 There are two ways to interpret the quantization condition \eqref{eq:quant}. First, it is useful to
 rewrite it by recalling
 that the Compton wavelength of a particle of rest energy $E_0$ is $\lambda_c = \frac{2\pi \hbar c}{E_0}$.
 Thus we have
 \begin{equation}\label{eq:quant-ii}
 r = \frac{2n+1}{4} \lambda_c.
 \end{equation}
The first interpretation of this  equation is that it demands
 that different fermions have different retardation
 parameters, given by \eqref{eq:quant-ii}. We might expect this to become problematic
 when, say, electrons interact with protons, neutrons, or neutrinos.
 A second interpretation declares that there is a universal
 retardation time in Nature, denoted $r$. In this note we will adopt the second
 point of view.

The question arises as to the magnitude of $r$ and $n$. The hypothesis we
are entertaining is that the modified Dirac equation \eqref{eq:mod-Dir}   should
apply to stable fermions whose propagation in vacuum would ordinarily be
described (to good approximation) by a standard Dirac equation.
Thus we are led to consider protons, neutrons, electrons,
and neutrinos.   The Compton wavelength of the
electron is $\lambda_c^e \sim 10^{-12}$m,  while for the proton and neutron we have
$\lambda^{p,n}_c/\lambda_c^e \sim 10^{-3}$, whilst the lightest neutrino probably has
$10^5 < \frac{\lambda^{\nu}_c}{\lambda_c^e} < 10^9$ \cite{PDG,Vogel:2008zzb}. Equation \eqref{eq:quant-ii} shows that $r$ is
bounded below by $\lambda_c/4$. Optimistically taking the smallest nonzero neutrino mass
 $\sim 1 eV$ we have
\begin{equation}\label{eq:r-bound}
r \succsim 10^{-5} {\rm cm}.
\end{equation}
This scale is uncomfortably large and we hence take the corresponding integer
for the lightest neutrino to be
of order one.

 Returning to \eqref{eq:ener-q} we have:
\begin{equation}\label{eq:mc2-dev}
E_0=mc^2+(-1)^n kc,
\end{equation}
This is not a deviation from Einstein's formula relating rest energy to mass,
 but simply the relation of the
inertial rest energy to the parameters of the modified Dirac equation.

Of course, the modified Dirac equations will lead to deviations from standard
physical results and hence we expect $k$ and $r$ to be small. One
interesting deviation is in the equivalence principle. The modified
 Dirac equation can be derived from an action principle
 \begin{equation}\label{eq:action}
 \int \vol \bar \psi \left( i \hbar \ID_+ - mc \right) \psi
 \end{equation}
 from which one can derive the energy-momentum tensor $T_{\mu\nu}$. For our purposes it will
 suffice to consider the action in a weak
 gravitational field
 \begin{equation}\label{eq:weak-grav}
 ds^2 = - (1-2 \Phi) dt^2 + dx^i dx^i,
 \end{equation}
  where $\Phi$ is the Newtonian
 gravitational potential, and extract the coefficient of $\Phi$ to obtain the gravitational
 rest energy. In the curved metric \eqref{eq:weak-grav} the Dirac
 operator $D = \gamma\cdot \p + S + \cdots $ with $S = \Phi \gamma^0 \p_0 + \half \gamma_i \p_i \Phi$
 Making this substitution, and using on-shell spinor wavefunctions for a particle
 at rest we find
\begin{equation}\label{eq:equiv-dev}
T_{00}=E_0\left(1- \mu (-1)^n\right)
\end{equation}
where
\begin{equation}\label{eq:def-mu}
\mu:= \frac{kr}{\hbar}
\end{equation}
 is a dimensionless number. We thus find a
 deviation from the equivalence principle
 \begin{equation}
 \frac{ T_{00} - E_0}{E_0} =(-1)^{n+1} \mu.
 \end{equation}
  Consequently, the parameter $\mu$ will have to be very small (less than about $10^{-13}$) not to contradict observational evidence \cite{Will:2010uh}.   Since $r$ is uncomfortably large  according to
  \eqref{eq:r-bound} it must be that $k/\hbar$ is small, something which will prove to be
  interesting in Section \ref{sec:Cosmo-Const} below. Indeed
  \begin{equation}\label{eq:k-upper}
  \frac{k}{\hbar} < 10^5 \mu {\rm cm}^{-1} < 10^{-8 } {\rm cm}^{-1} .
  \end{equation}
  The parameter $\mu$
   is a fundamental   constant of our ``theory" and links together the two key dimensionful parameters $r$, the shift, and $k$, the coefficient that measures the magnitude of the term that shifts the Dirac operator.  Moreover \eqref{eq:def-mu} involves Planck's constant, reflecting the quantum character of the parameter $kr$.  All these parameters are embodied in our basic choice of the shifted Dirac operator.

We close this section with three sets of remarks.

\subsection{Remarks regarding the formal modification of the Dirac operator}

\begin{enumerate}

\item  Let us comment on the modified Dirac equation for the other choices $\ID_-, \ID_s, \ID_c$.
Using $\ID_-$ instead of $\ID_+$ we obtain the same results by replacing particles with antiparticles.
For the modified Dirac equations using $\ID_c$ and $\ID_s$
 there is no  quantization condition such as \eqref{eq:quant}, whilst \eqref{eq:mc2-dev} is
 replaced by a transcendental relation between $E_0, m, k, r$. The equation with $\ID_s$ and
 $m=0$ is compatible with a massless dispersion relation but that with $\ID_c$ is not.  The violation of the
 equivalence principle \eqref{eq:equiv-dev} is similar in all four cases.

\item One may ask how to define the shifted version of the Klein-Gordon equation for
scalar fields. The
usual relation $D^2 = - \nabla^2$ is more complicated for modified Dirac operators
\eqref{eq:mod-Dir}. One simple possibility is to replace $D^2 \to \ID_+ \ID_-$. The
latter can be interpreted as a power series in $D^2$:
\begin{equation}
\ID_+ \ID_- = D^2 - 2\mu D^2 \left( \frac{\sinh r D}{rD}\right) - \frac{k^2}{\hbar^2}
\end{equation}
Replacing $D^2 \to - \nabla^2$ in this expression produces a candidate modified Klein-Gordon
operator.

 \item   We can extend, or couple, the Dirac operator to other fields.  Thus coupling to spinors again we get the
 operator $d+\delta$ acting on all differential forms (where $\delta$ is the Hodge adjoint of $d$).  We can exponentiate this operator but since it does not preserve the degrees of forms, we cannot just restrict it say to 2-forms.  Treating Maxwell's equations requires more care and we will come back to this later in Section
 \ref{ref:E-M}.

\item  Following Dirac, our search for relativistically
 shifted equations inevitably led us to spinors. An interesting question
 is how such shifted equations interact with supersymmetry. Perhaps one
 should shift field equations in superspace.

\end{enumerate}

\subsection{Further remarks on advanced and delayed equations in general }

\begin{enumerate}

\item Returning to  \eqref{eq:ret}, we note that it is an unusual equation since it involves
 a sum of a skew adjoint
 operator with a unitary operator. Thus it involves an element of an
 ``affine operator group.''  This group
    is in turn a degenerate form  of a semi-simple group, reminiscent of the
Wigner contraction of   semi-simple groups to inhomogeneous orthogonal groups
such as the  Poinar\'e group.

\item
Shifting a differential equation in a way that involves both retarded and advanced terms drastically alters its nature, even in the simple 1-dimensional case when  \eqref{eq:ret} is replaced by
\begin{equation}\label{eq:adv-ret-x}
\dot{x}(t)+k\{x(t-r)+x(t+r)\}=0.
\end{equation}
 This is no longer a simple evolution equation
 and the theory of such equations has hardly been developed.
 However, once again, there is an infinite-dimensional space of
  exponential solutions $x(t)= e^{-zt/r}$ where $z$ now satisfies the equation
\begin{equation}\label{eq:tran-2}
z = \mu (e^z + e^{-z}).
\end{equation}
As before, there is an infinite set of roots $z_n$ which tend to infinity for $n \to \pm \infty$.
An infinite-dimensional space of solutions is given by the linear combinations
\begin{equation*}
\sum_n x_n \exp(-z_n t/r).
\end{equation*}

\item
The roots of \eqref{eq:tran-2} always have nonzero real parts so once again there is
unacceptable behavior in the past or future. However, it should be noted that in the
limit $\mu \to \infty$ these real parts tend to zero. This is particularly
clear if one takes $k\to \infty$ in \eqref{eq:adv-ret-x}, since the solutions
plainly become anti-periodic with period $2r$. Thus, for large   $\mu$ there is a
finite-dimensional space of solutions with acceptable oscillatory behavior, at least in a
restricted time domain. Of course our interest is in the opposite case of very small $k$, but this might conceivably be related by some kind of duality. The parallel interchange of small and large values of $r$ arises in Fourier theory. In fact Feynman in \cite{FeynmanThesis} studies an equation similar to
 \eqref{eq:adv-ret-x} where $r$ represents the distance between a particle and a virtual image.
  He shows that,   surprisingly, there is a conserved energy. (In fact his system is equivalent to a standard physical one without retardation.) For Feynman $r$ is large while for us $r$ is small.

\item For our shifted Dirac operator we expect that  mathematical
difficulties of the sort encountered in the previous two remarks arise from the mixing of the positive
and negative frequencies in the presence of external forces.
In standard quantum field theory such mixing of states   forces the introduction
of Fock space in quantum field theory.  In other words our shifted Dirac
operator may be easy to define but its mathematical and physical implications
can be profound.  A link to quantum theory would not therefore be too surprising.

\end{enumerate}

\subsection{Some more phenomenological remarks}

\begin{enumerate}

\item We have not attempted to investigate the modified Dirac equation in nontrivial
electromagnetic backgrounds. Moreover, we have not attempted to include the effects
of quantum interactions and thus explore the effects on the standard successes of
QED such as Bhaba scattering, Compton scattering, the Lamb shift, the anomalous magnetic
moment etc.  This would
be a logical next step if our present speculations are to be taken further. The
existence of nontrivially interacting string field theories gives some hope that such
interactions can be sensibly included.

\item Since the modified Dirac equation \eqref{eq:mod-Dir} is not compatible with massless fermions it is not
at all obvious how to include important properties of the standard model such as chiral
representations of the gauge group and the Higgs mechanism into our framework.

\item It follows from \eqref{eq:quant-ii} that
 all particles have a Compton wavelength (and hence
 mass) related by a ratio of two odd integers. Of course any
 real number admits such an approximation, but  the numerator
 and or denominator in such an approximation cannot become
 too large without making $r$ too large. For the electron
 and proton these integers must be large, but this is not 
 obviously so for neutrinos. Perhaps this idea
 can be tested experimentally as our knowledge of neutrino
 masses improves.

\end{enumerate}

\section{The Shifted Einstein Equations}\label{sec:Einstein}

\subsection{Definition}\label{sec:Shift-Ricci}

Even if we do not know how shifted equations may relate to quantum mechanics we can still ask if there is any way of producing a shifted version of the Einstein equation of General Relativity.

\medskip

 We begin by recalling some standard differential geometry, initially in the
positive-definite Riemannian version. Recall there  
 are two natural second order differential operators 
 active on the space of 1-forms, both generalizing the Laplacian of flat space,

\begin{description}
\item[(i)] the Hodge Laplacian $= (d+\delta)^2=\Delta$
\item[(ii)] the Bochner Laplacian $= \nabla^*\nabla$ where
\end{description}

\begin{equation*}
\nabla : \Omega^1\rightarrow \Omega^1 \otimes \Omega^1
\end{equation*}

\noindent is the covariant derivative.

\noindent The Weitzenbock formula asserts that

\begin{equation}\label{eq:H-B}
\textrm{Hodge - Bochner = Ricci}
\end{equation}

\noindent This, on a compact manifold, was originally used by Bochner to prove that, if the Ricci tensor was positive definite, then there were no harmonic 1-forms and so the first Betti number was zero.

  Formula \eqref{eq:H-B} continues to hold for a Lorentzian manifold, though sign conventions have to be carefully checked. We replace $\nabla^*\nabla$ with $-\nabla^2$ and define $\delta$ using the Hodge star.
   More fundamentally, we view $D= d+\delta$ as a Dirac operator coupled to the  spin bundle. Then, we have
   \begin{equation}\label{eq:Mink-H-B}
   D^2 + \nabla^2 = \textrm{Ricci }
   \end{equation}
   as an operator equation on $1$-forms. We are indebted to Ben Mares for his careful calculations to confirm signs.

  Equation \eqref{eq:Mink-H-B} says that  the Einstein vacuum equations, Ricci $=0$, asserts the equality of the two Laplacians on 1-forms.  This indicates how we might define a shifted Hodge Laplacian   on 1-forms: We replace $D$ by one of the modified Dirac operators \eqref{eq:D-mod}. There are 10 ways to do this, but we will focus on only three: %
 \begin{subequations}\label{eq:Ricc-mod}
 \begin{align}
 \textbf{Ricci}_{++} := & c(\ID_+)^2 + \nabla^2 \label{eq:Ric-pp} \\
  \textbf{Ricci}_{+-} := & \ID_+ \ID_- + \nabla^2 \label{eq:Ricc-pm}\\
 \textbf{Ricci}_{ss} : = &  \ID_s^2  + \nabla^2 \label{eq:Ricc-ss}
 \end{align}
 \end{subequations}

In equation \eqref{eq:Ric-pp} we cannot  simply square the operator $\ID_+$
 because the resulting operator does not preserve the
  degree of forms. However, we can then compress it   on 1-forms by taking the composite operator
\begin{equation}\label{eq:Comp}
c(\ID_+^2) := P \left(\ID_+ \right)^2 I
\end{equation}
 where \textbf{$I$} is the inclusion $\Omega^1 \rightarrow \Omega^*$ and $P$ is the projection $\Omega^* \rightarrow \Omega^1$. The other two operators do not need compression.

The shifted Einstein equations should then be given by equating \textbf{Ricci} to zero. Written out,
these are the equations
 \begin{subequations}\label{eq:Ein-mod}
 \begin{align}
  \textrm{Ricci}-\frac{2k}{\hbar} D \sinh(rD) + \frac{k^2}{\hbar^2}\cosh(2rD)  & = 0  \label{eq:Ein-pp} \\
   \textrm{Ricci}-\frac{2k}{\hbar} D \sinh(rD) - \frac{k^2}{\hbar^2}  & = 0  \label{eq:Ein-pm} \\
    \textrm{Ricci}+\frac{2k}{\hbar} D \sinh(rD) + \frac{k^2}{\hbar^2}\sinh^2(rD)  & = 0  \label{eq:Ein-ss}
 \end{align}
 \end{subequations}

It is  not at all clear how to interpret these operator equations. We will comment further
on this point in Section  \ref{subsec:Einst-interp} below.
One interpretation is to regard
the operator as an expansion in $D^2$, then, using \eqref{eq:Mink-H-B}, convert this to
 an expansion in $\nabla^2$. Then
we can view higher order terms in $\nabla^2$ as an expansion in low energy and small momenta.

\subsection{The cosmological constant}\label{sec:Cosmo-Const}

If we adopt the viewpoint that the shifted Einstein equations \eqref{eq:Ricc-mod}
can be understood as an expansion in low energies then
 the leading    term is a well-defined equation on the metric given by
 replacing $D^2 \to {\rm Ricci}$ in \eqref{eq:Ein-mod}. Using the fact that
 $\mu$ must be small the modified Einstein equations become approximately
 \begin{subequations}\label{eq:Cosmo-Const}
 \begin{align}
  \textrm{Ricci}   & = -\frac{k^2}{\hbar^2}   \label{eq:Cosmo-const-pp} \\
  \textrm{Ricci}   & = \frac{k^2}{\hbar^2}   \label{eq:Cosmo-const-pm} \\
  \textrm{Ricci}   & = 0   \label{eq:Cosmo-const-ss}
 \end{align}
 \end{subequations}
    In other words the first correction to the Einstein equation produced by our shifted operators is to introduce a cosmological constant $\Lambda = \pm \frac{k^2}{\hbar^2}$. This relates our
     parameter $k$ to cosmology. Evidently, the theory can accommodate all three cases of positive, negative,
    and zero cosmological constant, and hence   the correct sign of the cosmological constant is hardly a triumph for our approach. The observational data support the case \eqref{eq:Cosmo-const-pm}.
    Any logical consideration which distinguishes or rules out some  of the possible shifted Ricci
    tensors would  be most interesting.

    The magnitude of $\Lambda$ is also interesting. As we have seen,
    observed bounds on violations of the equivalence principle
    imply that $k/\hbar$ is small and hence $\Lambda$ is small. In fact, the observational
      evidence
  \footnote{See, for example, \cite{Tegmark:2003ud}.}   gives an order of magnitude for $\Lambda$
  corresponding to an energy density $\rho_{vac} \sim (1 eV)^4$ and
  hence $\Lambda = \frac{8 \pi G}{c^4} \rho_{vac}$ of order
\begin{equation}
\Lambda \sim 10^{-56}\textrm{cm}^{-2}
\end{equation}
 and hence
\begin{equation}\label{eq:mu-mag}
\frac{k}{\hbar} \sim 10^{-28}\textrm{cm}^{-1},
\end{equation}
much smaller than the upper bound \eqref{eq:k-upper}.  If we now take the order of magnitude of $r$ derived from our quantization condition
then it follows from \eqref{eq:r-bound} that
\begin{equation}\label{eq:mu-fin}
\mu \succsim  10^{-33}.
\end{equation}

\subsection{Further comments on interpretation}\label{subsec:Einst-interp}

In Section \ref{sec:Shift-Ricci} we showed formally how to shift the Ricci tensor, giving formula
\eqref{eq:Ricc-mod}, and we focused on the lowest order correction which gave the cosmological constant.  We then noted the order of magnitudes that emerged.  Notably the estimate \eqref{eq:mu-fin} for our dimensionless parameter $\mu=kr/\hbar$.  But, as we noted, it is not at all clear how to interpret the operator equation (acting on 1-forms)
\eqref{eq:Ein-mod}.  We will now discuss this question.

  In the first place, although $D$ does not preserve the degrees of forms,   a power series in the dimensionless quantity $r^2 D^2=r^2 \Delta$  should make sense (even non-perturbatively) on the space of 1-forms.

  Second it seems reasonable to restrict \eqref{eq:Ein-mod} to 1-forms which are locally solutions of the wave equation
\begin{equation}\label{eq:wave}
- \nabla^2 \phi=0
\end{equation}
and hence determined by initial conditions along a space-like 3-space.  In view of  \eqref{eq:H-B}, on such 1-forms $\phi$, we have

\begin{equation*}
\Delta \phi = \textrm{Ricci}\ (\phi)
\end{equation*}

\noindent so that $\Delta =D^2$ acts tensorially on such $\phi$.  Unfortunately it does not preserve the solutions of \eqref{eq:wave}, since $\Delta$ need not commute with the Ricci operator.  However if we are looking for a perturbation expansion in powers of $\mu$ this procedure might be made to work.

 An alternative idea is to regard the equations \eqref{eq:Ein-mod} as equations on an
infinite-dimensional Hilbert space, namely $\Omega^1(M_4)$. That is we consider the equation
\begin{equation}\label{eq:alt}
( \ID_+ \ID_- + \nabla^2)\phi=0
\end{equation}
as an equation for the pair $(g,\phi)$ where $g$ is the metric tensor and $\phi$ is a 1-form.
 \footnote{This is reminiscent of Einstein \ae ther theory. See, for example,
 \cite{ArkaniHamed:2002sp}. However, a minimal requirement would appear to be that
the  restriction to any $x\in M_4$ of the $\phi$'s solving  \eqref{eq:alt} should span $T_x^*M_4$. } 
 This needs detailed investigation and again it could first 
 be looked at perturbatively in powers of $\mu$.  
 However it should be emphasized that \eqref{eq:alt}  makes sense non-perturbatively.

Finally, we remark that in string field theory one very naturally
runs into exponentials $\exp[ \alpha' \nabla^2] $ acting on fields. This is usually not
considered to be too disturbing since at energy-momenta small compared to the string
scale such terms are close to $1$, and at energy-momenta on the order of the string
scale the usual notion of  commutative space and time might be breaking down in any case.
Certainly, topology changing effects in string theory take place at that scale.
Witten has in the past speculated that string theory would necessarily lead to a revision
of quantum mechanics. (His main reason being that there is no dilaton-field
in 11-dimensional supergravity.)  There is thus some remote connection here to Witten's
speculation.

\section{The Shifted Einstein-Maxwell Equations}\label{ref:E-M}

  In addition to shifting the Dirac operator and the Ricci operator we should also shift the Maxwell operator

\begin{equation}\label{eq:max-op}
d^*F_A=d^*dA.
\end{equation}
One way to do this is to  use Kaluza-Klein reduction and the shifted Einstein equations
 in five-dimensional spacetime.  That is, we take the 5-dimensional space $M_5$ to be a
principal circle bundle over the Lorentzian spacetime $M_4$ equipped with  a connection
$\Theta$ and a metric:
\begin{equation}\label{eq:KK-ansatz}
ds^2 = e^{2\sigma} \Theta^2 + d\bar s^2
\end{equation}
where $d\bar s^2$ is the pullback of a metric on the four-manifold $M_4$
 and $\sigma$ is the dilaton field, again pulled back from $M_4$.  The shifted Einstein equations for
 such a metric produce the shifted Einstein-Maxwell equations in four dimensions.

 In addition to the shifted Einstein-Maxwell equations the
  equation of motion for the scalar $\sigma$ is amusing.
 When $F_A=0$ it is simply
\begin{equation}\label{eq:Radius}
- \bar\nabla^2 e^{\sigma} = \epsilon {k^2 \over \hbar^2} e^{\sigma}
\end{equation}
where $\epsilon=\pm 1$,
so the radius $e^{\sigma}$ of the KK circle is an eigenfunction of
the Laplacian on the four-dimensional spacetime. If we take
the metric to be deSitter space, and consider spatially homogeneous
solutions to \eqref{eq:Radius} then the general solution is
\begin{equation}\label{eq:gensol}
e^{\sigma}  = a_+ e^{ \sqrt{\epsilon} {kc \over \hbar} t} + a_- e^{ -\sqrt{\epsilon}{kc\over \hbar} t}
\end{equation}
Thus, in general the   circle radius grows exponentially in the past or the future.
This looks potentially devastating since the circle starts
growing or shrinking exponentially on a time-scale
$\sim {\hbar  \over k c}$. However,  if we take
$k/\hbar \sim 10^{-28} cm^{-1}$ then  this timescale
is approximately $(3\pi)^{-1} \times 10^{11}$ years. For comparison
the age of the universe is approximately $1.4 \times 10^{10}$ years!
  The observation that ${1\over \sqrt{\Lambda} c}$ is the same order of magnitude
as the age of the universe is a version of the famous ``coincidence problem'' in
modern cosmology. It means that we just happen to live at an era in
the history of the universe when the effects of the cosmological constant
change from being negligible to being dominant. In our discussion the
coincidence ``problem'' turns out to be a blessing, not a curse.
\footnote{Still, the existence of a massless scalar with gravitational
strength couplings is problematic in view of fifth-force experiments \cite{Will:2010uh}.}  Of course, once
the exponential behavior
starts to become important one should surely no longer use the
zero-mode approximation (i.e. neglecting the $\CO(D^2)$ corrections).

We close with three remarks

\begin{enumerate}

\item    Clearly the above procedure could be extended to include nonabelian Yang-Mills equations
by considering Kaluza-Klein reduction on spaces with nonabelian isometry groups.

\item  This discussion shows that the idea
of shifting geometric operators is a natural process, akin to geometric quantization.  It would be fascinating to find any relation between the two processes especially since our shifting process includes the gravitational field, a target not yet achieved in quantization.

\item  The fact that our treatment rests fundamentally on spinors, the Dirac operator, and Kaluza-Klein
 theory suggests possible connections to super-gravity and string theory.

\end{enumerate}

\section{Conclusion}

In this paper we have, very tentatively, put forward a speculative new idea that seems worth exploring.  The idea is to introduce in a natural geometric way operators which shift (i.e. retard or advance) the basic operators of mathematical physics.  This includes the Dirac, Maxwell and Ricci operators (occurring in the Einstein equations of GR).  The shifting involves just two key physical parameters

\enlargethispage{1cm}

\begin{equation} \label{eq:r-mag}
r \succsim 10^{-5} {\rm cm}
\end{equation}
\begin{equation}\label{eq:k-mag}
\frac{k}{\hbar}  \sim 10^{-28} {\rm cm}^{-1}.
\end{equation}
Here $r$ measures the timeshift and $k$ measures the magnitude of the shift. There is a natural quantization condition
 \begin{equation}\label{eq:quant-ii-con}
 r = \frac{2n+1}{4} \lambda_c
 \end{equation}
where $\lambda_c$ is the Compton wavelength of a stable fermion. Of course, this has a quantum-mechanical
aspect and involves Planck's constant $\hbar$.

  The constant $k$ is at first sight arbitrary (except that it clearly must be very small).  However once we introduce our shifted Ricci operator we find that $k^2/\hbar^2$ is related to the cosmological constant.  Using the observed value of this gives the estimate \eqref{eq:k-mag}.

 Thus the two key constants $r,k$ are determined by physical observations at the atomic and cosmological scales respectively.  This is a satisfactory situation. It is also reminiscent of some of the
ideas of T. Banks \cite{Banks:2010tj}.

Note that our approach has a dimensionless fundamental constant linking $r,k$ and $\hbar$

\begin{equation}
\mu=kr/\hbar \hspace{20pt}\mu \succsim  10^{-33}.
\end{equation}

We leave the reader with three key questions

\begin{enumerate}

\item What is the correct interpretation of equation \eqref{eq:Ein-mod}?

\item Can   the above ideas can be made into a coherent model of physics, compatible
with the  successes of the Standard Models of particle physics and modern cosmology ?

\item How is this idea of shifted equations related to quantum mechanics?
We leave this to others and to the future.  There are tantalizing hints of possible connections,
not least the philosophical and mathematical difficulties on both sides!

\end{enumerate}

\begin{flushright}
M.F. Atiyah,\\
School of Mathematics\\
JCMB, The King's Buildings\\
Mayfield Road\\
Edinburgh,      EH9 3JZ\\
Scotland\\
UK
\end{flushright}

\smallskip
\thispagestyle{empty}

\begin{flushright}
G.W. Moore\\
NHETC\\
Dept. of Physics and Astronomy\\
Rutgers, The State University of New Jersey\\
136 Frelinghuysen Road\\
Piscataway, NJ 08854-8019 \\
USA
\end{flushright}

\end{document}